\documentclass[aps,prl,floats,floatfix, twocolumn,amssymb,amsmath,
superscriptaddress,nofootinbib,showpacs,longbibliography]{revtex4-2}

\usepackage[T1]{fontenc}
\usepackage[utf8]{inputenc}
\usepackage{txfonts} 
\usepackage[normalem]{ulem} 

\usepackage{graphicx}
\usepackage[linktocpage,breaklinks]{hyperref}
\usepackage[capitalize]{cleveref}
\usepackage[usenames,dvipsnames]{xcolor}
\hypersetup{colorlinks=true,citecolor=NavyBlue,
linkcolor=NavyBlue,urlcolor=NavyBlue}

\usepackage{multirow,array}
\DeclareMathAlphabet{\pazocal}{OMS}{zplm}{m}{n}
\usepackage{microtype}
\usepackage{subfigure}

\usepackage{journals}
\usepackage{upgreek}
\usepackage{stmaryrd} 

\newcommand{\nn}{\nonumber\\}
\newcommand{\dd}{{\rm d}}
\newcommand{\pa}{\partial}
\newcommand{\scal}{\tilde{\varphi}}

\newcommand{\uiuc}{\affiliation{Department of Physics and Illinois Center for Advanced Studies of the Universe,\\University of Illinois Urbana-Champaign, Urbana, Illinois 61801, USA}}
\newcommand{\aei}{\affiliation{Max Planck Institute for Gravitational Physics (Albert Einstein Institute), D-14476 Potsdam, Germany}}
\newcommand{\igc}{\affiliation{Institute for Gravitation and the Cosmos, The Pennsylvania State University, University Park, PA 16802, USA}}
\newcommand{\psu}{\affiliation{Department of Physics, The Pennsylvania State University, University Park, PA 16802, USA}}

\begin{document}
\title{Donutization Inside Neutron Stars: Shell-Localized Scalar Fields}

\date{\today}

\author{Hao-Jui Kuan}
\email{hjkuan@illinois.edu}
\uiuc

\author{Alan Tsz-Lok Lam}
\igc\psu\aei

\author{Jacquelyn Noronha-Hostler}
\uiuc

\author{Nicol{\'a}s Yunes}
\uiuc

\begin{abstract}
Heavy scalar fields ($m_\phi\gtrsim10^{-9}$ eV) in scalar-tensor gravity are expected to be hidden from neutron-star observations because their Compton wavelength is sub-stellar.  
We show that neutron stars can nevertheless scalarize by forming a shell-localized profile, suppressed at their center and exterior but peaked in their interior.
This \emph{donutization} reshapes the effective equation of state, making hadronic stars mimic quark-star mass-radius behavior or hybrid-star behavior with split stable branches, and breaks the $I$--$Q$ relation, while remaining hidden from binary pulsar observations.
\end{abstract}
\maketitle

\noindent
\textit{Introduction.}
Scalar-tensor extensions of general relativity provide a canonical framework~\cite{Brans:1961sx} for exploring strong-field deviations from Einstein's theory, with neutron stars serving as natural laboratories.
Their most remarkable signature is that a nonperturbative instability can be activated to endow neutron stars with ``scalar charge'' (a process known as scalarization~\cite{Damour:1993hw,Damour:1996ke,Doneva:2022ewd}), while passing Solar-System and cosmological tests~\cite{Anderson:2016aoi,dePireySaintAlby:2017lwc}.
As gravitational-wave~\cite{Zhao:2019suc,Xie:2024xex,Gupta:2025utd} and binary-pulsar observations~\cite{Freire:2012mg,Antoniadis:2013pzd,Shao:2017gwu,Anderson:2019eay,Zhao:2022vig} tighten bounds on light scalars, viable scalar-tensor phenomenology is pushed toward massive fields with Compton wavelengths shorter than binary scales~\cite{Yazadjiev:2016pcb,Staykov:2018hhc,Xu:2020vbs,Hu:2021tyw,Popchev:2018fwu,Danchev:2020zwn,Morisaki:2017nit,Rosca-Mead:2020bzt,Degollado:2024oyo}.

The prevailing picture, however, is that scalarization simply switches off once the reduced Compton wavelength $\lambdabar = 1/m_\phi$ falls much below the typical stellar radius.
Fields with $m_\phi\agt10^{-9}$ eV ($\lambdabar\alt0.15$~km) are generally believed to be too short-ranged to support scalarization at all, yielding a theory that is observationally and phenomenologically indistinguishable from general relativity~\cite{Ramazanoglu:2016kul,Doneva:2022ewd}.
This picture assumes that scalarization produces a centrally-peaked profile that decreases monotonically toward the stellar surface, which, as we show here, is actually unnecessary.
We demonstrate that neutron stars can scalarize even for $m_\phi\gg10^{-9}$~eV, with the scalar field localizing into a finite-radius interior shell.
We call this shell-localized scalarization \emph{donutization}: in three dimensions, the support is shell-like, while in meridional cross-section it appears annular, like a donut viewed from the top.

The mechanism is simple.
After the scalar equation is written as a sourced massive Klein--Gordon equation, the scalar field can be represented as an integral over the effective source with a Yukawa-type Green's kernel.
Unlike the massless Green's function, this kernel decays exponentially with the separation between the field point and each source element, making the scalar response local on scales of order $\lambdabar$.
At the same time, the effective matter source for scalarization can be suppressed deep inside the core of the star.
The scalar field consequently localizes where the source remains active, forming a finite-width interior shell.
What makes this profile distinctive is its sharpness.
The outer edge can span sub-kilometer scales, producing a steep scalar gradient that significantly softens the effective equation of state in the outer core and modifies the matter distribution.

Donutization is different from the usual scalarization phenomenology in massless or light-scalar theories.
There, scalarization is typically understood as the growth of a scalar profile that fills the star and produces a long-range scalar field~\cite{Damour:1993hw,Damour:1996ke,Doneva:2022ewd}.
Here, the large scalar mass changes the problem qualitatively: the exterior field is screened, the interior field very close to the core can also be suppressed, and the scalar stress is concentrated inside a shell.
The important effect is thus not simply that the star carries scalar hair, but that the shell reorganizes the effective pressure support of otherwise hadronic matter.

This reorganization can imitate two familiar dense-matter phenomena.
When donutization suppresses the effective pressure at low density, the scalarized sequence can acquire self-bound-like scaling, $M\!\propto\!R^3$, resembling quark-star mass-radius curves~\cite{Alcock:1986hz,Rahimi:2024rip,Balkin:2023xtr}.
When the same mechanism amplifies a soft region of the hadronic equation of state in the astrophysical mass range, the sequence can instead split into two stable branches separated by unstable configurations, resembling mass twins from a first-order phase transition~\cite{Gerlach:1968zz,Glendenning:1998ag,Schertler:2000xq,Alford:2013aca,Benic:2014jia,Christian:2018jyd,Pang:2020ilf,Tan:2021ahl}.
We verify this stability structure with numerical relativity simulations.

We use massive Damour--Esposito-Far{\`e}se (DEF) theory~\cite{Damour:1992we,Damour:1993hw,Damour:1996ke} as a concrete setting, but the physics is not special to this model.
In the high-$m_\phi$ regime, the short Compton wavelength suppresses binary-scale dipole radiation and makes dynamical scalarization during inspiral difficult to trigger~\cite{Kuan:2023hrh}, so the strongest existing binary constraints do not directly exclude the effect in the parameter range studied here.
At the same time, the shell-localized interior field can still produce large structural effects: the anisotropic stress of the donutized shell can make a hadronic star resemble a self-bound quark star~\cite{Alcock:1986hz}, break the equation-of-state insensitive relation between moment of inertia and quadrupole moment ($I$--$Q$ relation~\cite{Yagi:2013awa,Yagi:2013bca,Pani:2014jra}), and split the stable equilibrium sequence into disconnected branches separated by configurations that we verify are dynamically unstable with fully relativistic evolutions.
Taken together, our findings transform the high-$m_\phi$ regime of scalar-tensor gravity from a presumed null result into a phenomenology with calculable signatures.

\vspace{0.2cm}
\noindent
\textit{Basic Equations.}
For definiteness, we work in the massive DEF model, defined in the Jordan frame by
\begin{align}
    S = \frac{1}{16\pi} \int \mathrm{d}^4x \, \sqrt{-g^{\rm J}} 
    \left[ \phi \mathcal{R} - \frac{\omega(\phi)}{\phi} 
    \nabla_{\mu} \phi \nabla^{\mu} \phi - U(\phi) \right] 
    + S_{\mathrm{matter}} \,,
\end{align}
where $g^{\rm J}$ and $\mathcal{R}$ denote the determinant and Ricci scalar of the metric $g_{\mu \nu}^{\rm J}$, with Greek letters in index lists representing spacetime coordinates, $[\omega(\phi)+3/2]^{-1}=B\ln\phi$ with $B$ a constant that controls the coupling of the scalar field to matter, and $U=2m_\phi^2\phi^2\scal^2/B$ the Jordan-frame potential with $\scal^2=2\ln\phi$~\cite{Shibata:2013pra,Taniguchi:2014fqa,Kuroda:2023zbz,Kuan:2023hrh,Lam:2025jsk,Lam:2024azd}.
Although the Jordan frame makes the physical metric explicit, the Einstein frame simplifies the gravitational sector by making the scalar minimally coupled to the metric; the nonminimal coupling is instead shifted into an explicit scalar-matter coupling.
Using the Einstein-frame metric $g_{ab}$, related to the Jordan-frame metric by
$g_{\mu \nu}^{\rm J}=e^{-\scal^2/2}g_{\mu \nu}$, and defining
$\varphi:=\scal/\sqrt{B}$, the field equations become
\begin{align}\label{eq:Gmunu}
    G_{\mu \nu} = 8\pi T_{\mu \nu}^{\rm tot} =  8\pi T_{\mu \nu} + 
    8 \pi T_{\mu \nu}^{\rm s}\,,
\end{align}
where the Jordan-frame baryonic stress-energy tensor and potential map to the Einstein frame as $T_{\mu \nu} = e^{-\scal^2/2} T^{\rm b,J}_{\mu \nu}$ and $V=U/(4 \phi^2) = m_\phi^2\scal^2/(2B)=m_\phi^2\varphi^2/2$, while the scalar-field stress energy tensor is
\begin{align} \label{eq:Tab-scalar}
    T_{\mu \nu}^{\rm s} &= \frac{1}{4\pi} \left[(\pa_{\mu}\varphi) (\pa_{\nu}\varphi) - \frac{1}{2} g_{\mu \nu}\left(\pa_{\alpha}\varphi\right)\left(\pa^{\alpha}\varphi\right)
    - g_{\mu \nu} V(\varphi) \right]
\end{align}

We construct stationary equilibria with the Komatsu–Eriguchi–Hachisu method~\cite{Komatsu:1989zz,Cook:1992} (see~\cite{Doneva:2013qva,Doneva:2016xmf,Doneva:2018ouu,Staykov:2023ose,M:2026whf} for similar setups in massless and relatively-light $m_\phi$ cases), which adopts quasi-isotropic coordinates~\cite{Bonazzola:1993zz,Shibata:2007zzb,Paschalidis:2016vmz} for the line element, given by
\begin{align}
    \dd s^2\!\!=\!\!-e^{\upgamma+\uprho}\dd t^2 
    \!\!+\!\! e^{\upgamma-\uprho}r^2 \sin^2\theta \left(\dd \phi-\omega\dd t\right)^2
    \!\!+\!\! e^{2\alpha}\left(\dd r^2+r^2\dd\theta^2\right)\,.
\end{align}
The field equations can be found in~\cite{Doneva:2013qva}, but we rewrite the scalar field equation here as
\begin{align}\label{eq:KG}
    \left(\triangle_2-m_\phi^2\right) \left(\varphi e^{\upgamma/2} \right) &=
    \varphi  \triangle_2( e^{\upgamma/2}) \nn 
    &+ \left[ 2\pi B Te^{2\alpha} + m_\phi^2
    \left(e^{2\alpha}-1\right) \right] \varphi e^{\upgamma/2} 
\end{align}
with 
$\triangle_2:=\pa_r^2 + (2/r) {\pa_r} + 
[({1-x^2})/{r^2}]\pa_x^2 
- (2 {x}/{r^2})\pa_x$  and $x=\cos\theta$.
The related Yukawa-like Green's function~\cite{Jackson62} involves the modified spherical Bessel functions of the first ($i_\ell$) and second ($k_\ell$) kinds and reads
\begin{align}
    G(r,r')=\frac{m_\phi}{4\pi}\sum_{\ell=0}^{\infty}\left(2\ell+1\right)i_\ell\left(m_\phi r_<\right) k_\ell\left(m_\phi r_>\right)
    P_{\ell}\left(\hat{r}\cdot\hat{r'}\right)\,,
\end{align}
where $r_>=\max(r,\,r')$, $r_<=\min(r,\,r')$ and $P_{\ell}$ denotes Legendre polynomials~\cite{note1}.
The modified spherical Bessel functions are exponentials in disguise: at large argument, $i_\ell(z)$ grows like $e^z/(2z)$ while $k_\ell(z)$ decays like $\pi e^{-z}/(2z)$, so each partial wave carries the same short-range Yukawa suppression.
Imposing equatorial reflection symmetry, we then obtain
\begin{align}\label{eq:varphi}
    \varphi(r,x)\!&=\!-m_\phi \!\!\sum_{\substack{\ell=0\\ \ell\ {\rm even}}}^{\infty}
    \!\!(2\ell+1)P_{\ell}(x)\,\Bigg[\!\!\int_0^{r} \!\!\!\! {\rm d}r'r'^2 f(r') 
    i_\ell\left(m_\phi r'\right)
    k_\ell\left(m_\phi r\right)\nn
    &+
    \int_r^{\infty} {\rm d}r'r'^2 f(r') 
    i_\ell\left(m_\phi r\right)
    k_\ell\left(m_\phi r'\right)
    \Bigg] \; e^{-\gamma/2}
\end{align}
for $f(r')=\int_0^1{\rm d}x'P_{\ell}(x') S_\varphi(r',x')$.
We also adopt the same grid structure as~\cite{Cook:1992,Cook:1993qj,Cook:1993qr,Stergioulas:1994ea}, where the semi-infinite interval $r\in[0,\infty)$ is mapped to the definite interval $s\in[0,1)$ via $s=r/(r+r_{\rm eq})$, with $r_{\rm eq}$ the equatorial radius at equilibrium.
During the iterations toward convergence, the grid self-adapts to the neutron star size.

\vspace{0.2cm}
\noindent
\textit{Scalarized interior.}
We model baryonic matter in the Jordan frame as a perfect fluid, $T^{\rm b,J}_{\mu \nu} = \rho_{\rm b} \, h_{\rm b} \, u_{\mu} u_{\nu} + p_{\rm b} \, g_{\mu \nu}^{\rm J}$, with rest-mass density $\rho_{\rm b}$, relativistic specific enthalpy $h_{\rm b}=\left(\varepsilon_b+p_b\right)/\rho_{\rm b}$, baryonic energy density $\varepsilon_{\rm b}$, baryonic pressure $p_{\rm b}$, and fluid $4$-velocity $u^{\mu}$.
However, the kinetic term of scalar field in the scalar-field stress-energy tensor of Eq.~\eqref{eq:Tab-scalar} can lead to anisotropic pressure.
Given this,  it is useful to interpret the total stress energy tensor, $T_{\mu \nu}^{\rm tot}$, as an effective anisotropic fluid one, with some effective energy density and effective radial and tangential pressures. For the nonrotating configurations used to characterize the effective
radial response, these quantities can be read from the mixed diagonal
components,
\begin{align}
\varepsilon_{\rm eff}
&:= - T^t{}_t{}^{\rm tot}
= A^4\varepsilon_{\rm b}+\varepsilon_{\rm s}\,,\\
p_{\rm eff}^{(r)}
&:= T^r{}_r{}^{\rm tot}
= A^4p_{\rm b}+p_{\rm s}^{(r)}\,,\\
p_{\rm eff}^{(t)}
&:= T^{\theta}{}_{\theta}{}^{\rm tot}
= T^{\phi}{}_{\phi}{}^{\rm tot}
= A^4p_{\rm b}+p_{\rm s}^{(t)}\,,
\end{align}
where for a spherically-symmetric scalar, the scalar-field contributions to these effective quantities are 
\begin{align}\label{eq:peff}
    \varepsilon_{\rm s} =\frac{(\pa_r\varphi)(\pa^r\varphi) + m_\phi^2\varphi^2}{8\pi} = -p^{(t)}_{\rm s}
    \quad
    p^{(r)}_{\rm s} &=\frac{(\pa_r\varphi)(\pa^r\varphi)- m_\phi^2\varphi^2}{8\pi}\,.
\end{align}
The scalar field gradients increase $\varepsilon_{\rm s}$ and $p_{\rm s}^{(r)}$, but the mass term increase  $\varepsilon_{\rm s}$, while it decreases $p_{\rm s}^{(r)}$. Therefore, $\varepsilon_s$ and $-p_s^{(t)}$ are positive definite, while $p_s^{(r)}$ can change sign; the latter's sign and magnitude determine whether scalarization stiffens or softens the baryonic equation of state.  This sign change is absent in the massless limit, because then the scalar contribution to $\varepsilon_{\rm eff}$ and $p_{\rm eff}^{(r)}$ are positive definite, although the tangential pressure remains negative. Thus, the low-density softening discussed below is not a generic feature of scalarization, but a consequence of the massive potential term overwhelming the scalar gradient term in the shell-localized regime.

\begin{figure}[tb]
    \centering
    \includegraphics[width=\columnwidth]{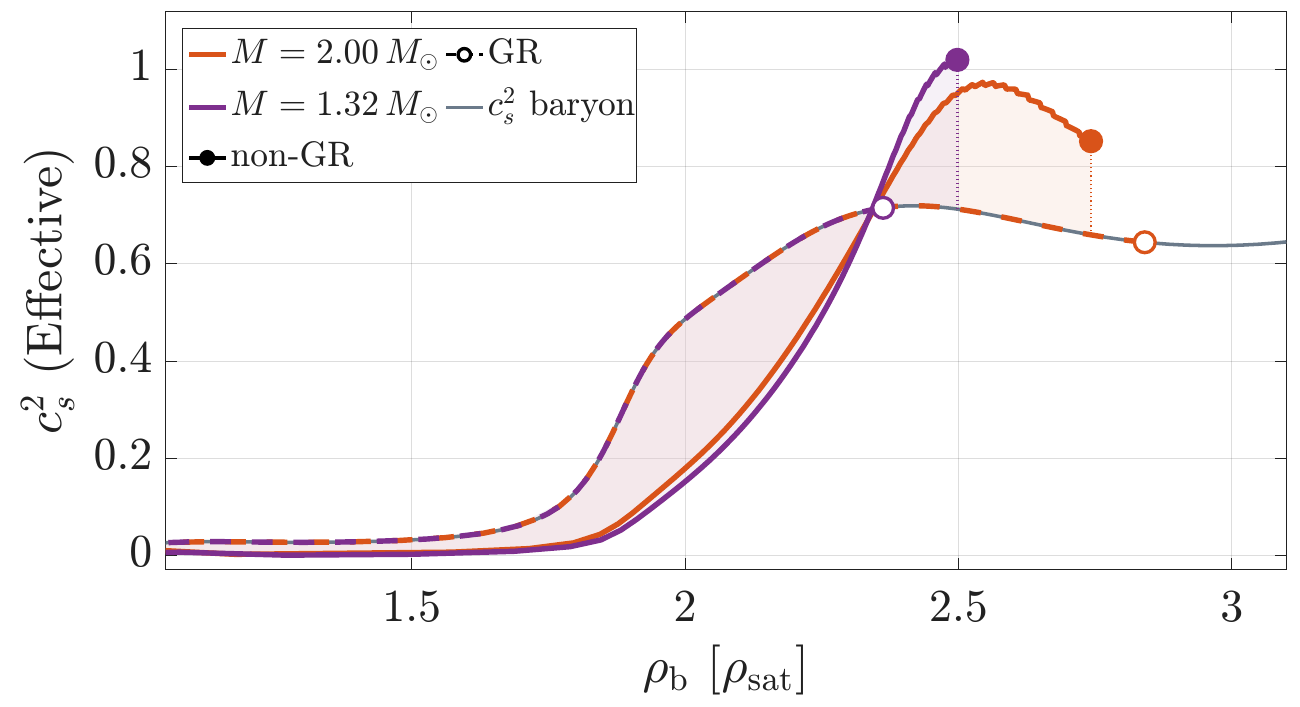}
    \caption{Square of effective speed of sound as a function of baryon rest-mass density in units of nuclear saturation density $\rho_{\rm sat}=2.7\times10^{14}\,{\rm g\,cm^{-3}}$.
    Thick curves show the total effective $c_s^2({\rm Effective})$ of \cref{eq:cs2}, while the gray curve gives the baryonic baseline $c_s^2({\rm Baryon})$. 
    The offset from the gray curve encodes the scalar contribution $c_{s}^2({\rm Scalar})$, and is depicted by the shaded area.
    Filled and open circles denote the central values of the DEF and general-relativistic neutron stars, respectively, with $M=1.32\,M_\odot$ in purple and $M=2.00\,M_\odot$ in red.
    Results are shown for $m_\phi=1.3\times10^{-10}\,{\rm eV}$ and $B=400$.}
    \label{fig:cs}
\end{figure}

To illustrate how this anisotropy reshapes stellar interiors, we use an equation of state containing $npe\mu$ matter based on a relativistic mean-field Lagrangian, identical to the IU-FSU model~\cite{Fattoyev:2010mx}, with the seven coupling constants specified in~\cite{Legred:2025aar}. 
\Cref{fig:cs} shows the baryonic speed of sound squared, $c_s^2({\rm Baryon})=\dd p_{\rm b}/\dd\varepsilon_{\rm b}$, and the scalar contribution, $c_s^2({\rm Scalar})=\dd p_{\rm s}/\dd\varepsilon_{\rm b}$, as functions of the baryon rest-mass density, where our effective speed of sound squared is then
\begin{equation}\label{eq:cs2}
    c_s^2({\rm Effective}) := \frac{\dd p_{\rm eff}}{\dd\varepsilon_{\rm b}}
    = c_s^2({\rm Baryon})+c_s^2({\rm Scalar}) .
\end{equation}
While $c_s^2({\rm Baryon})$ must obey the usual causality and stability constraints, $c_s^2({\rm Scalar})$ and  $c_s^2({\rm Effective})$ do not.
For a given baryonic equation of state, $c_s^2({\rm Baryon})(\rho_{\rm b})$ is the same for all stars on a $M-R$ sequence but $c_s^2({\rm Scalar})(\rho_{\rm b})$ depends on the underlying $\rho_{\rm b}(r)$ profile inside the star, such that $c_s^2({\rm Effective})(\rho_{\rm b})$ also varies.
In other words, scalarized stars do not all have the same \emph{effective} equation of state, and thus, \cref{fig:cs} shows two separate curves with different scalar field contributions $c_s^2({\rm Scalar})$ that depend on the central density of the star.

An important effect occurs when the baryonic sound speed has a peak near the density where the scalar contribution changes sign.
For the IU-FSU example in \cref{fig:cs}, this happens at $\rho_{\rm b}^{\rm peak}\sim2.4\rho_{\rm sat}$.
Below this density, $c_s^2({\rm Scalar})<0$, so the scalar field softens the effective response; above it, $c_s^2({\rm Scalar})>0$, so the scalar field rapidly stiffens the effective response.
The scalar field therefore \textit{amplifies} an existing feature of the baryonic equation of state rather than introducing a completely independent density scale.

This sign change explains how scalarization moves stars through the $\rho_b$ range within the equation of state.
For a general-relativistic sequence, a stellar mass $M_1$ corresponds to a central density $\rho_{{\rm b},1}^{\rm c}$, while the scalarized configuration with the same mass generally has a different central density, $\hat{\rho}_{{\rm b},1}^{\rm c}$.
If the scalarized star lies mostly below $\rho_{\rm b}^{\rm peak}$, the scalar contribution softens the interior and the star reaches a larger central density, $\hat{\rho}_{{\rm b},1}^{\rm c}>\rho_{{\rm b},1}^{\rm c}$.
If the scalarized star lies above $\rho_{\rm b}^{\rm peak}$, the scalar contribution stiffens the core and the same mass can be supported at a smaller central density, $\hat{\rho}_{{\rm b},1}^{\rm c}<\rho_{{\rm b},1}^{\rm c}$.
Near the transition, these effects can compensate, making the intersection of the scalarized and general-relativistic $M$--$R$ curves only weakly sensitive to the coupling parameter $B$; see also Fig.~4 of~\cite{Doneva:2022ewd}.

The resulting effective response is very soft at low $\rho_{\rm b}$ but rises rapidly between $2$ and $2.5\rho_{\rm sat}$, resembling the behavior used to support ultraheavy neutron stars~\cite{Tan:2020ics,Tan:2021ahl} or quarkyonic matter~\cite{McLerran:2018hbz}.
The key difference is that, here, the baryonic equation of state is fixed.
Different scalarized stellar models sample different density profiles and therefore acquire different effective responses, so the scalar field can generate new behavior in the $M$--$R$ curve without replacing the underlying hadronic equation of state.

\begin{figure}[tb]
    \centering
    \includegraphics[width=\columnwidth]{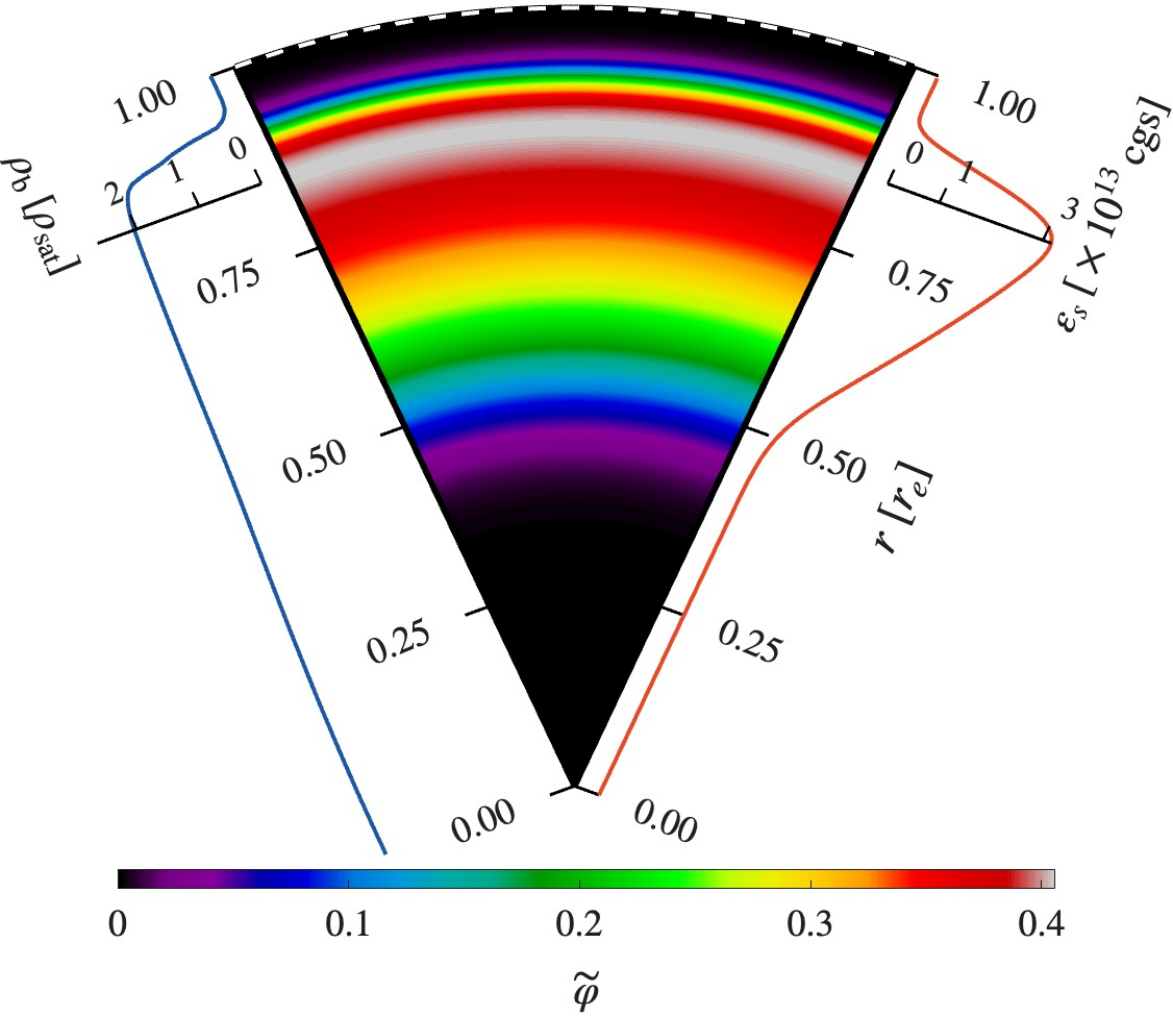}
    \caption{Schematic diagram of a donutized neutron star interior. Focus on the annular structure of the scalar field, which is shown both as a color map and through the scalar field energy density on the right side. Compare this to the baryon rest-mass density, shown on the left side, as well as the radial coordinate in units of the equatorial radius of the star on the right side. The neutron star considered here has $M=1.8\,M_\odot$, $m_\phi=1.3\times10^{-9}\,{\rm eV}$ and $B=2.5\times10^4$. 
    }
    \label{fig:pizza}
\end{figure}

\vspace{0.2cm}
\noindent
\textit{Donutization.}
For $m_\phi>4\times10^{-10}$ eV ($\lambdabar\alt0.5$~km), Yukawa suppression confines the scalar field to the stellar interior and exponentially suppresses its exterior tail.
For sufficiently massive stars, pressure effects can also quench scalarization in the inner core. Defining the baryonic trace anomaly $\Theta_{\rm b}\equiv \varepsilon_{\rm b}-3p_{\rm b}=-T_{\rm b}$, Eq.~\eqref{eq:KG} shows that the matter term favors scalarization where $T_{\rm b}<0$, or equivalently where $\Theta_{\rm b}>0$, is sufficiently large~\cite{Podkowka:2018gib,Fujimoto:2022ohj,Marczenko:2022jhl,Mendes:2024fel,Ji:2025ati}. As the core stiffens and $\Theta_{\rm b}$ decreases toward zero or changes sign, the scalar instability weakens there.
The combined suppression near the surface and in the inner core localizes the scalar profile inside a finite-radius shell. In a meridional slice, this shell appears annular, as shown and explained in \cref{fig:pizza}; this visual structure motivates the name \emph{donutization}, since the meridional slice resembles a donut or torus when viewed from above. The three-dimensional support, however, is more accurately a spherical shell in the nonrotating limit, or an oblate shell for rotating stars.
The shell-like scalar configurations are also found in prior works~\cite{Rosca-Mead:2020bzt,Morisaki:2017nit}, while here we identify the high-mass neutron-star regime and connect it to quark-star or hybrid star mimicry, $I$--$Q$ violation, and branch splitting.

\begin{figure}
    \centering
    \includegraphics[width=\columnwidth]{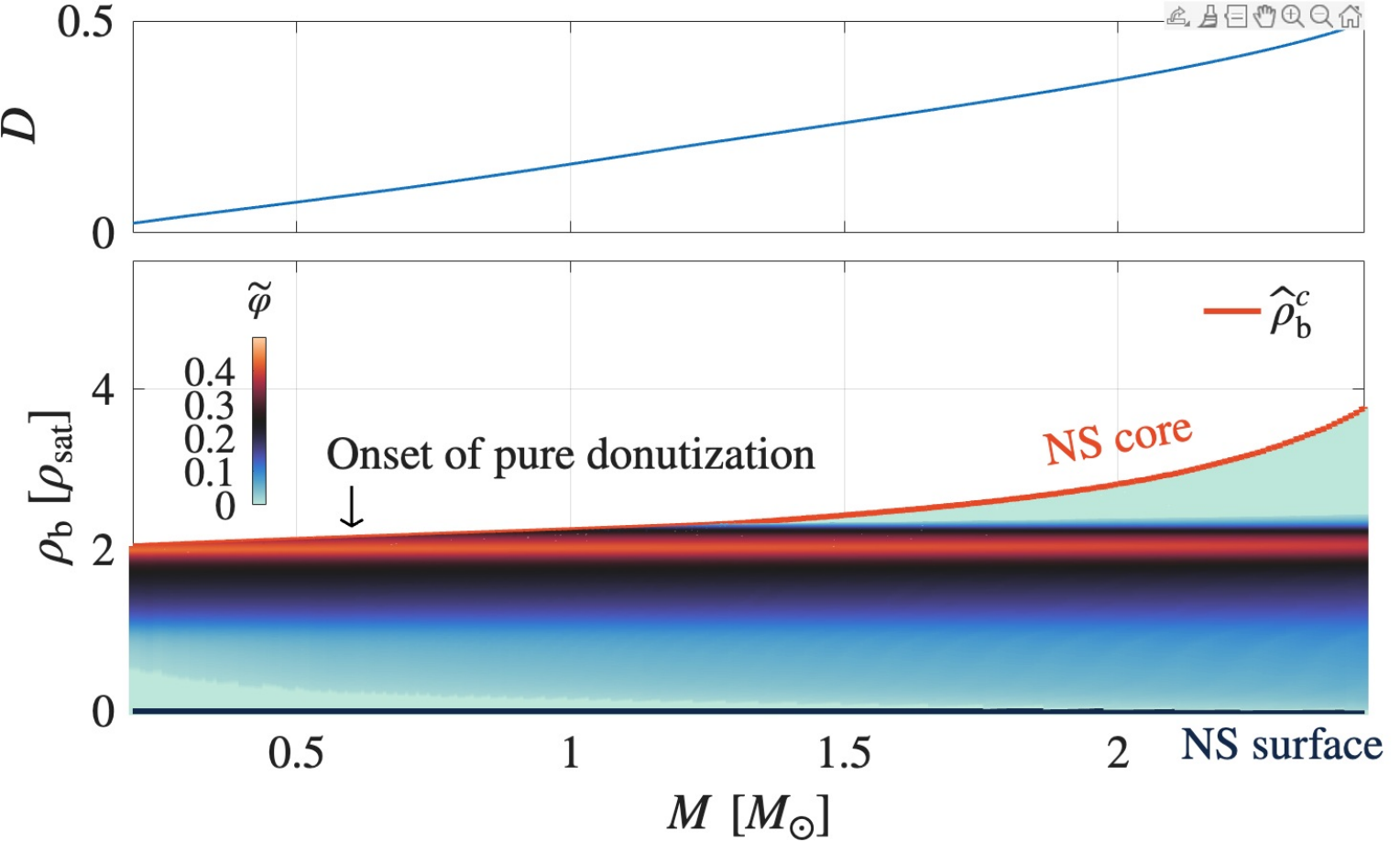}
    \caption{Donut number versus gravitational mass (top) and Jordan-frame scalar amplitude versus baryon mass density (bottom), for a sequence of nonrotating neutron stars.
    Each vertical stripe is one equilibrium configuration at fixed gravitational mass: moving downward along a stripe traces the star from its central density to the surface, where $\rho_{\rm b}=0$.
    At larger masses, the scalar support is expelled from the center and forms an off-center annulus; correspondingly, $D$ grows rapidly once the shell is fully developed for $M\gtrsim0.6\,M_\odot$.
    DEF parameters are $m_\phi=1.3\times10^{-9}$~eV and $B=2.5\times10^4$.
    }
    \label{fig:map10}
\end{figure}

\begin{figure*}
    \centering
    \includegraphics[width=1\columnwidth]{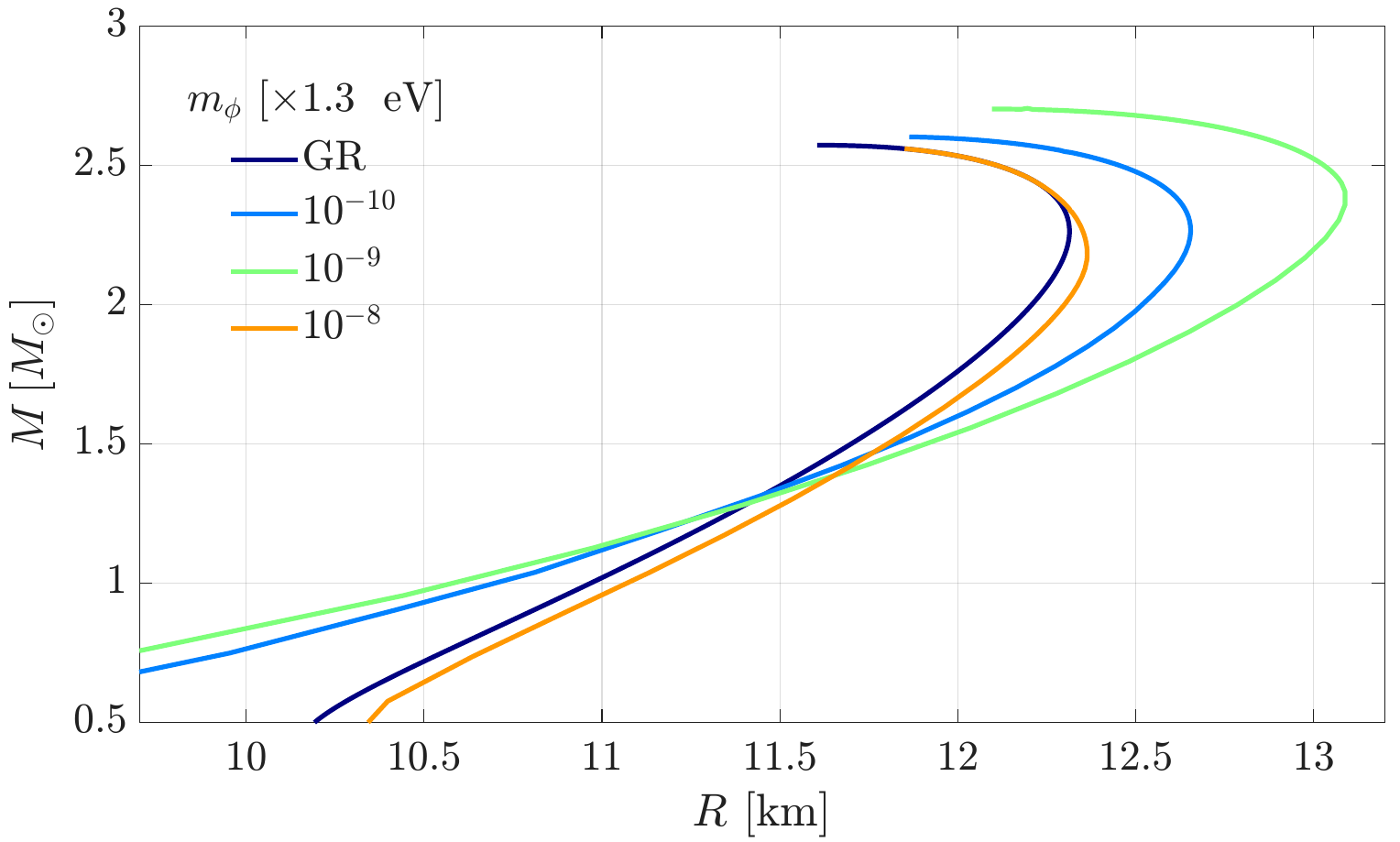}
    \includegraphics[width=\columnwidth]{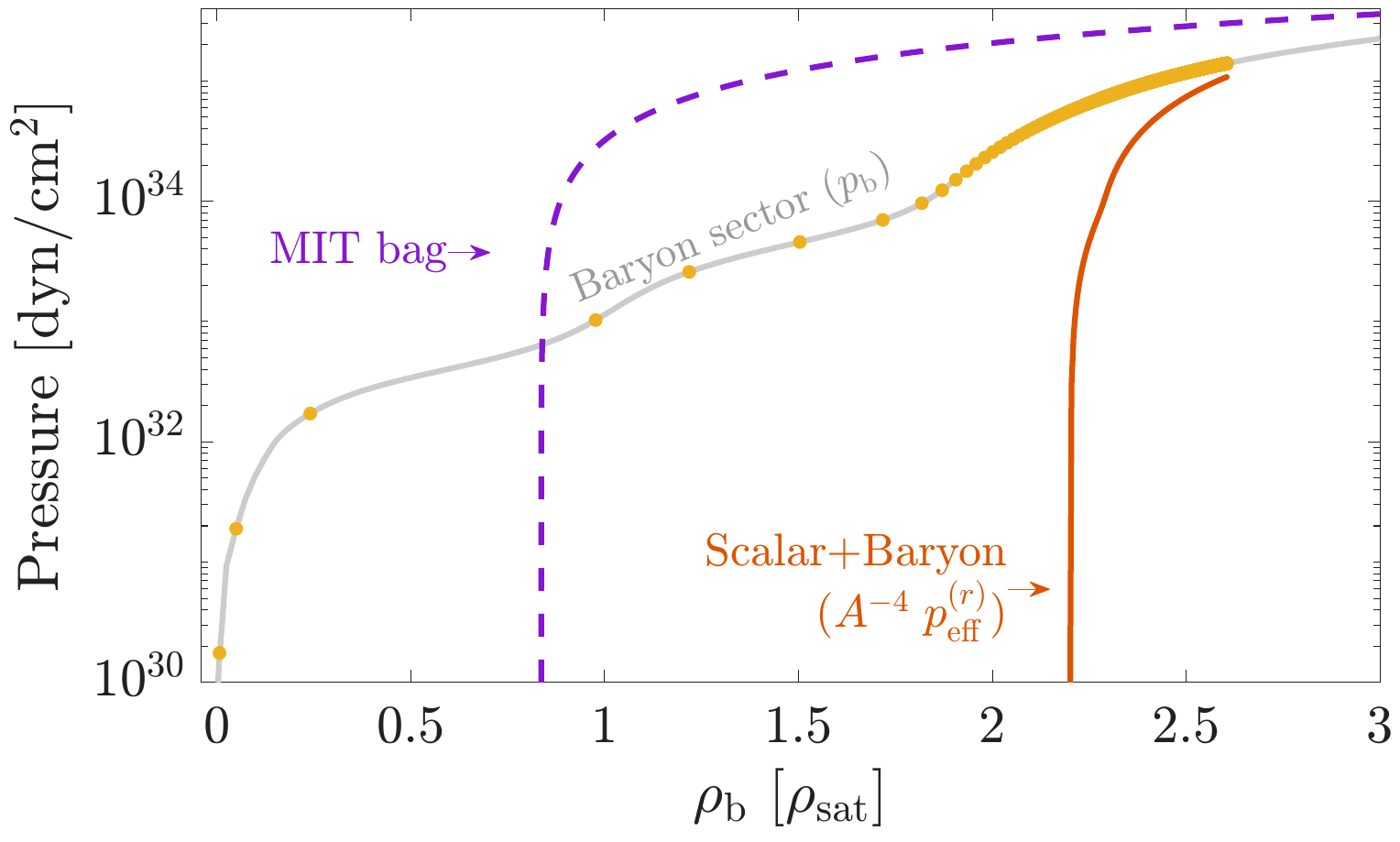}
    \caption{\textbf{Left:} Equilibrium, static neutron stars with the IU-FSU equation of state in DEF theories and various scalar masses. 
    The coupling strengths $B$ have been chosen so that the $M$--$R$ curves are close to general relativity when $M\simeq1.4\,M_\odot$; in particular, $B=400,~5\times10^4,$ and $7\times10^6$ for the lower to higher $m_\phi$.
    \textbf{Right:}
    Meridional cross-section of a shell-localized scalar field inside a neutron star pertaining to the modified PS baryonic equation of state (gray). Also shown are the baryonic sector of a scalarized neutron star (yellow dots), the effective total pressure including the scalar-field contribution (red), and a representative MIT bag-model quark-star equation of state (purple). The yellow dots and red curve are extracted from a scalarized neutron star with $M=1.8\,M_\odot$. The model corresponds to the green line on the left.
    }
    \label{fig4}
\end{figure*}

To quantify the transition from ordinary scalarization to donutization, we introduce a geometric donut number $D$ constructed from superlevel sets of the Jordan-frame scalar amplitude.
This diagnostic weights the scalar amplitude by the baryonic mass per unit shell thickness, making it sensitive to both the strength of scalarization and the geometry of the scalar support; the formal definition is given in the End Matter.
Centrally peaked profiles at $M\alt0.6M_\odot$ give only moderate values of $D$, whereas $D$ grows once the scalar develops a persistent off-center peak and shell-like support across many contour levels, as tracked in~\cref{fig:map10}.

\vspace{0.2cm}
\noindent
\textit{Quark-Star Mimicry.}
At particular values of $(m_\phi,B)$, the mass term in
\cref{eq:peff} dominates over the gradient term inside the
shell-localized region, driving $p_{\rm s}^{(r)}<0$.
The scalar stress then suppresses the effective radial pressure
at finite, intermediate baryon density, lower than the central
density but still well inside the star, as indicated by the red
annulus in \cref{fig:pizza}.
The right panel of \cref{fig4} shows the resulting effective
pressure-density relation: the underlying matter equation of state
remains hadronic, but the scalar contribution reshapes
$p_{\rm eff}^{(r)}(\rho_{\rm b})$ so that the pressure nearly
vanishes at $\rho_{\rm b}\sim2.2\rho_{\rm sat}$.
This finite-density pressure onset is reminiscent of self-bound
quark-star matter, and in particular of MIT-bag-model equations
of state~\cite{Bodmer:1971we,Witten:1984rs,Chodos:1974je,Farhi:1984qu,Alcock:1986hz}.

The analogy to quark stars is phenomenological but still interesting.
We are not implying that scalarization produces quark matter.
Rather, the scalar field changes the effective pressure support
of an otherwise hadronic star.
The left panel of \cref{fig4} shows the corresponding global
response: above a minimum mass, the scalarized sequence follows
a quark-star-like $M$--$R$ trend (with a positive slope at low mass), even though the microscopic
equation of state has not changed.
Unlike self-bound quark stars, $M$ does not go to zero at
very small $\rho_{\rm b}^{\rm c}$, because this is below the regime
where scalarization appears. The $M$--$R$ sequence instead eventually bends to the right again, toward 
large $R$ for very low $\rho_{\rm b}^{\rm c}$ (see End Matter).

\vspace{0.2cm}
\noindent
\textit{Branch Splitting and Stability.}
The same shell-localized scalar stress that makes a hadronic star mimic a
quark-star sequence can also split the stable equilibrium branch.
To identify which parts of the resulting curve are dynamically accessible,
we must revisit the usual turning-point diagnostic.
For a one-parameter sequence of static neutron stars in general relativity,
the onset of radial instability is marked by the usual maximum-mass
turning point, at central density $\rho_{\rm b}^{\rm max}$, where
$\partial M/\partial\rho_{\rm b}^{\rm c}$ changes
sign~\cite{Sorkin:1981jc,Sorkin:1982ut,Friedman:1988er}.
This is the static limit of the turning-point theorem, which relates
stability changes to extrema along equilibrium sequences at fixed
conserved quantities~\cite{Sorkin:1981jc,Sorkin:1982ut,Friedman:1988er}.
In scalar-tensor gravity, the equilibrium space is enlarged by the scalar
degree of freedom. In massless DEF theory, the asymptotic scalar value
provides an additional parameter of this space~\cite{Damour:1992we,Harada:1998ge}.
In massive theories, the potential fixes the scalar to its asymptotic
minimum, so this boundary value is no longer freely specifiable.
We must therefore treat sign changes in
$\partial M/\partial\hat{\rho}_{\rm b}^{\rm c}$ as diagnostics, rather than
as a substitute for a dynamical stability check.
Related subtleties also arise in multi-scalar models with nontrivial
target-space geometry~\cite{Doneva:2019ltb,Kuan:2021yih}, where parts
of the scalarized branch can have larger $M$ at smaller
$\hat{\rho}_{\rm b}^{\rm c}$ and still be stable against radial perturbations.

Previous numerical evolutions with the DEF implementation~\cite{Lam:2025jsk}
of the \texttt{SACRA-2D} code~\cite{Lam:2025pmz} showed that the ordinary
maximum-mass point of the static scalarized sequence, at
$\hat{\rho}_{\rm b}^{\rm max}$, still marks a change of stability in massive
scalar-tensor theory.
In the high-$m_\phi$ regime studied here, however, the scalarized sequence
can develop an additional low-mass extremum before reaching
$\hat{\rho}_{\rm b}^{\rm max}$ for certain baryonic equations of state.
This produces a kink in the $M$--$R$ curve, which we call a
\emph{reverse turning point}.
It corresponds to a local minimum of the $M$--$\rho_{\rm b}^{\rm c}$ curve,
a feature recently argued to separate stable from unstable neutron-star
solutions~\cite{Balkin:2023xtr,Unluturk:2025zie,Muniz:2025egq,Staykov:2026ojk,Ozinan:2026bne}.

The physical interpretation of this reverse turning point is that the scalar field amplifies structure already present in the baryonic equation of state.
As shown in \cref{fig:cs}, the scalar contribution can soften density
intervals where the baryonic equation of state is already soft, while
stiffening denser regions.
A mild feature in the hadronic equation of state can then become a much
more pronounced feature in the $M$--$R$ sequence.
This makes equations of state with a soft outer core, or with a small
first-order or nearly first-order transition, especially useful diagnostic cases. 
For such baryonic equations of state, the scalar field can amplify the softening,
leading to a splitting of the $M$--$R$ curve and the emergence of twin-like
scalarized branches.

We demonstrate this enhancement with static configurations constructed
from a modified Pandharipande-Smith (PS) equation of state~\cite{Pandharipande:1975,Pandharipande:1976},
whose soft outer core makes it a useful diagnostic case.
The PS equation of state is purely hadronic, but includes pion condensation
and a small first-order phase transition when $\Delta$ baryons appear.
We use a piecewise-polytropic approximant of the PS equation of state and
raise the core adiabatic index from $\Gamma_3=2.365$ to $4$ in the notation
of~\cite{Read:2008iy}, so that the general-relativistic maximum mass exceeds
$2\,M_\odot$.
The PS approximant still contains the relevant phase-transition structure:
in general relativity it produces an inflection point in the $M$--$R$
sequence, but not mass twins.

\begin{figure*}
    \centering
    \includegraphics[width=\linewidth]{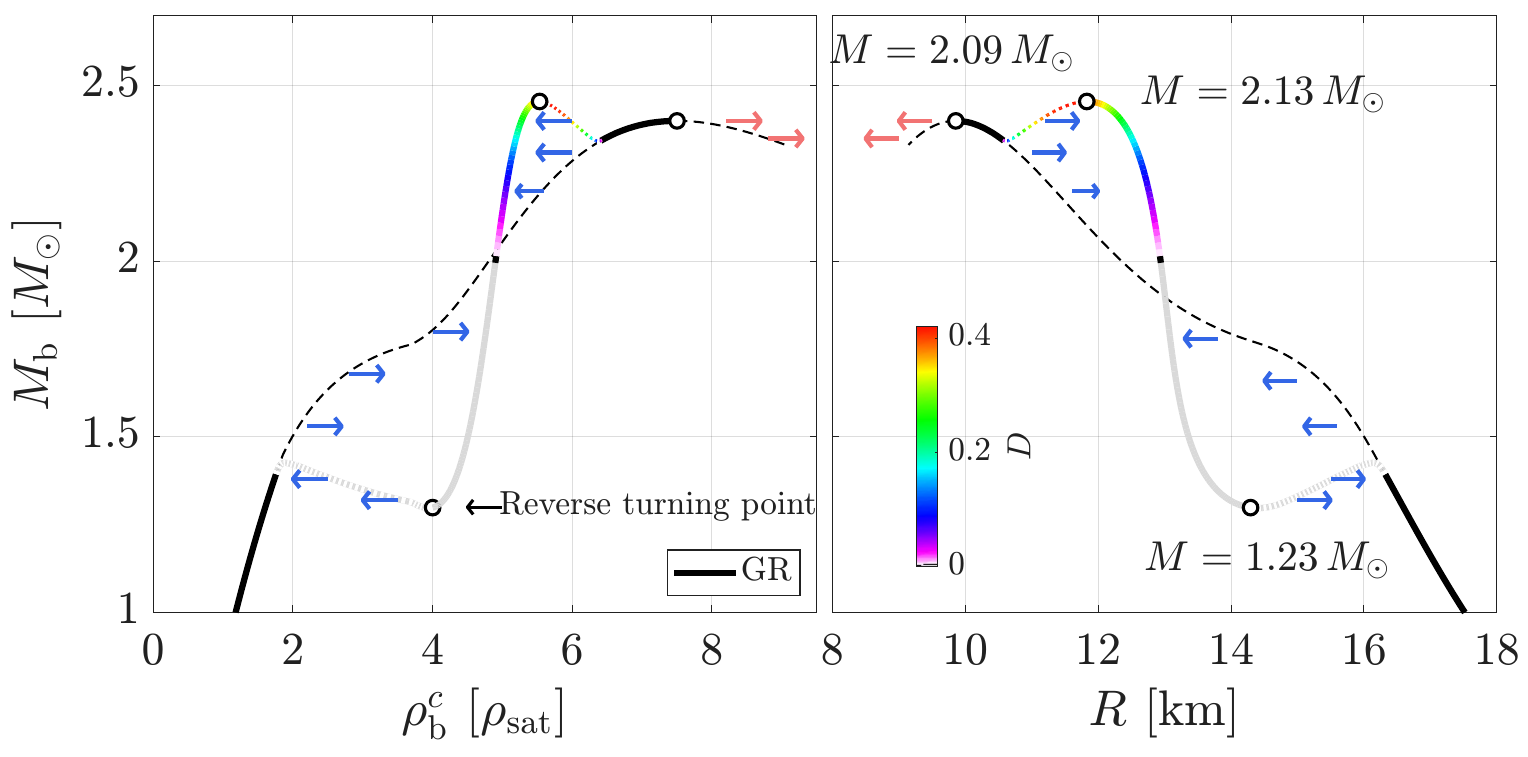}
    \caption{Baryonic mass versus central rest mass density (left) and stellar radius (right) for neutron star equilibria in the DEF model ($m_\phi = 1.3 \times 10^{-9}$~eV and $B = 2.5 \times 10^4$) with a modified PS equation of state.
    The general relativity branch is shown in black, while the scalarized branch is color-coded by the donut number $D$; the donut number is non-zero but small for scalar fields peaking at the neutron star center (gray).
    Stable and unstable segments are shown with solid and dashed lines respectively.
    Open circles mark turning points, with the standard ones at the maximum mass and the reverse one at the local minimum (see the text in the right panel), labeled by the corresponding $M$.
    Arrows indicate the non-linear migration direction of unstable equilibria: red arrows mark configurations that collapse to black holes, and blue arrows mark those that migrate to the lower-binding-energy branch.
}
    \label{fig6}
\end{figure*}

We then test the reverse turning point directly by constructing scalarized
and nonscalarized equilibria, evolving them in $1+1$ numerical relativity
(i.e.~in spherical symmetry), and tracking whether each model remains stable,
migrates along the same $M_{\rm b}$ sequence, or collapses to a black hole.
We also evolved a subset of models in axisymmetry, and found that our $2+1$ results 
are consistent with the $1+1$ evolutions.
\Cref{fig6} summarizes the stability classification, migration directions, and corresponding $M_{\rm b}$--$R$ relations.
As anticipated, the stable equilibria split into two disconnected branches separated by an unstable branch.
Near the reverse turning point, this structure is analogous to the twin-star branching produced by a strong first-order phase transition~\cite{Gerlach:1968zz,Glendenning:1998ag,Schertler:2000xq,Alford:2013aca,Benic:2014jia,Christian:2018jyd,Pang:2020ilf,Tan:2021ahl,second-split-note}.
Thus, the reverse turning point empirically marks a dynamical stability boundary: models on the intermediate segment are unstable and migrate toward one of the stable branches or collapse.

\vspace{0.2cm}
\noindent
\textit{The $I$-$Q$ relation.}
A confined scalar field can leave the $M-R$ relation nearly unchanged, while still modifying the equation of state insensitive relation between the moment of inertia $I$ and the quadrupole moment $Q$.
Although $I$ and $Q$ are extracted from the asymptotic properties of the metric, their interrelation depends on the stellar interior and on the approximate self-similarity of isodensity surfaces~\cite{Stein:2013ofa,Yagi:2014qua,Yagi:2016bkt}.
Scalar-induced pressure anisotropy disrupts this structure, in a manner similar to strongly magnetized neutron stars~\cite{Haskell:2013vha,Yagi:2015hda}.

The $I$--$Q$ relation was first identified in general relativity as an approximately equation-of-state-insensitive relation among bulk properties of neutron stars~\cite{Yagi:2013awa,Yagi:2013bca,Carson:2019rjx}.
Its robustness has since been tested in massless scalar--tensor theories~\cite{Pani:2014jra,Doneva:2014faa,Doneva:2016xmf,Popchev:2018fwu,Danchev:2020zwn,Hu:2021tyw} and in a broad class of modified-gravity models~\cite{Sham:2013cya,Kleihaus:2014lba,Doneva:2015hsa,Kleihaus:2016dui,Doneva:2017jop,Gupta:2017vsl}.
In related massive scalar-tensor models with relatively light scalars, $m_\phi=\mathcal{O}(10^{-11})$ eV, strongly scalarized stars are known to produce sizable deviations from the general-relativistic relation~\cite{Popchev:2018fwu,Danchev:2020zwn,Hu:2021tyw}.
Those deviations, however, occur in regions of coupling strength and scalar mass already tightly constrained by binary-pulsar and gravitational-wave observations~\cite{Freire:2012mg,Antoniadis:2013pzd,Shao:2017gwu,Anderson:2019eay,Zhao:2022vig,Zhao:2019suc,Xie:2024xex}, so the apparent breakdown does not constitute an independent practical limitation of the relation.

In this paper, our $m_\phi$ values are orders of magnitude larger than those previously studied, a regime where the shortened Compton wavelength suppresses binary-scale scalar radiation~\cite{Morisaki:2017nit,Degollado:2024oyo}.
For a representative theory with $\lambdabar\simeq0.15$ km, corresponding to $m_\phi=1.3\times10^{-9}$ eV, we consider five equations of state with a wide range of underlying assumptions and degrees-of-freedom, spanning radii of 11--14 km for a $1.35\,M_\odot$ star.
We specifically choose  $B=2.5 \times 10^4$ such that our $M-R$ relations are nearly identical to general relativity for this example.
As we show in the End Matter (see \cref{fig:IQ}), most of the $I$--$Q$ relation exhibits appreciable deviations from the general-relativistic equation-of-state-insensitive relation.
Compactness-based probes test scalar-tensor gravity through changes in the $M-R$ relation and, for pulse-profile measurements, through the photon propagation determined by the exterior geometry~\cite{Sotani:2017rrt,Silva:2018yxz,Silva:2019leq,Xu:2020vbs,Hu:2021tyw}.
The $I$--$Q$ relation supplies a complementary diagnostic, revealing deviations even when the $M$--$R$ curve is nearly unchanged.
This effect is therefore a practical caveat for using the $I$--$Q$ relation in model-independent tests.

\vspace{0.2cm}
\noindent
\textit{Discussion.}
Although a short Compton wavelength suppresses exterior scalar charges and binary-scale radiation, large scalar mass does \textit{not} simply erase scalar-tensor phenomenology.
Instead, donutization can confine the scalar field to a shell-localized state inside the star.
This exterior screening hides the field from observations most sensitive to long-range scalar effects, while the interior shell can enhance subtle features in the baryonic equation of state, making hadronic stars mimic quark stars or mass-twin branches.
The same anisotropic interior stress can also break the $I$--$Q$ relation, even for values of $(B,m_\phi)$ where the $M$--$R$ sequence is nearly identical to general relativity.

Donutization blurs the usual separation between equation of state inference and tests of gravity, because its effects would ordinarily be interpreted as consequences of dense-matter microphysics.
More broadly, neutron stars may act as strong-field detectors of hidden bosonic sectors that are invisible in weak-field or binary-scale tests.
The most direct connection is to scalar portals, dilatons, or moduli whose effective potentials depend on dense matter, but the same logic motivates extensions to axion-like sectors with electromagnetic or gravitational couplings.
In such theories, a shell-localized interior field could become a time-dependent source during mergers, potentially feeding scalar or axion radiation, axion--photon conversion, or parity-violating gravitational-wave signatures.
Thus, the natural next step is not only three-dimensional numerical relativity, but also coupling the donutized interior to conversion channels normally studied in magnetospheres or cosmology.

\vspace{0.2cm}
\noindent
\textit{Acknowledgment.---}
H.J.K and N.Y. acknowledge support from the Simons Foundation through Award No.~896696, the Simons Foundation International through Award No.~SFI-MPS-BH-00012593-01, and the NSF through Grant No.~PHY-25-12423.
A.T.L.L. acknowledges support from NASA under award No.~80NSSC25K7213.
J.N.H. acknowledges support from the US-DOE Nuclear
Science Grant No. DE-SC0023861.
Numerical computations were, in part, performed on the cluster Sakura at the Max Planck Computing and Data Facility, the cluster BinAC2 supported by the High Performance, and the Cloud Computing Group at the Zentrum f{\"u}r Datenverarbeitung of the University of T{\"u}bingen.

\bibliography{inspire,notinspire}

\clearpage
\appendix*
\section*{End Matter}
\section{S.1.~~~One-dimensional evolutions and stability test}
We perform spherical evolutions with the \texttt{SACRA-2D} infrastructure by imposing one-dimensional symmetry in the Cartoon method.
Together with additional implementations developed after~\cite{Lam:2025jsk}, this extended codebase is referred to as \texttt{SACRA-AEI}.
In a direct one-dimensional Cartoon implementation, one may introduce ghost layers in both the $\pm y$ and $\pm z$ directions.
Because of symmetry, however, all derivatives involving $z$ can be written in terms of derivatives with respect to $x$ and $y$.
We therefore introduce ghost layers only in the $\pm y$ direction.
For a scalar $Q$, vector $Q^i$, and tensor $Q^{ij}$, the nontrivial derivative identities are
\begin{align}
\begin{split}
    &\partial_{zz} Q = \partial_{yy} Q \\
    &\partial_{i} Q = 0 = \partial_{ij} Q
    \text{ otherwise except for $\partial_{x} Q$, $\partial_{xx} Q$}
\end{split}
\end{align}
\begin{align}
\begin{split}
    \partial_z Q^z &= \partial_y Q^y ,\quad
    \partial_{zz} Q^x = \partial_{yy} Q^x ,\quad
    \partial_{xz} Q^z = \partial_{xy} Q^y ,\\
    \partial_{j} Q^i &= 0 = \partial_{jk} Q^i
    \text{ otherwise except for $\partial_{x} Q^x$, $\partial_{xx} Q^{x}$}
\end{split}
\end{align}
\begin{align}
\begin{split}
    \partial_{x} Q^{zz} &= \partial_{x} Q^{yy} ,\quad
    \partial_{z} Q^{xz} = \partial_{y} Q^{xy} ,\\
    \partial_{xx} Q^{zz} &= \partial_{xx} Q^{yy} ,\quad
    \partial_{zz} Q^{xx} = \partial_{yy} Q^{xx} ,\quad
    \partial_{zz} Q^{yy} = \partial_{yy} Q^{zz} \\
    \partial_{xy} Q^{xy} &= \partial_{xz} Q^{xz} =
    \partial_{yy} Q^{xx} 
    + \frac{1}{2} \partial_{yy} Q^{yy}
    - \frac{3}{2} \partial_{yy} Q^{zz}\\
    \partial_{yz} Q^{yz} &= \frac{1}{2} \left( \partial_{yy} Q^{yy} - \partial_{yy} Q^{zz} \right)\\
    \partial_{k} Q^{ij} &= 0 = \partial_{kl} Q^{ij}
    \text{ otherwise except for $\partial_{x} Q^{xx}$, $\partial_{xx} Q^{xx}$}
\end{split}
\end{align}

\begin{figure}
    \centering
    \includegraphics[width=\columnwidth]{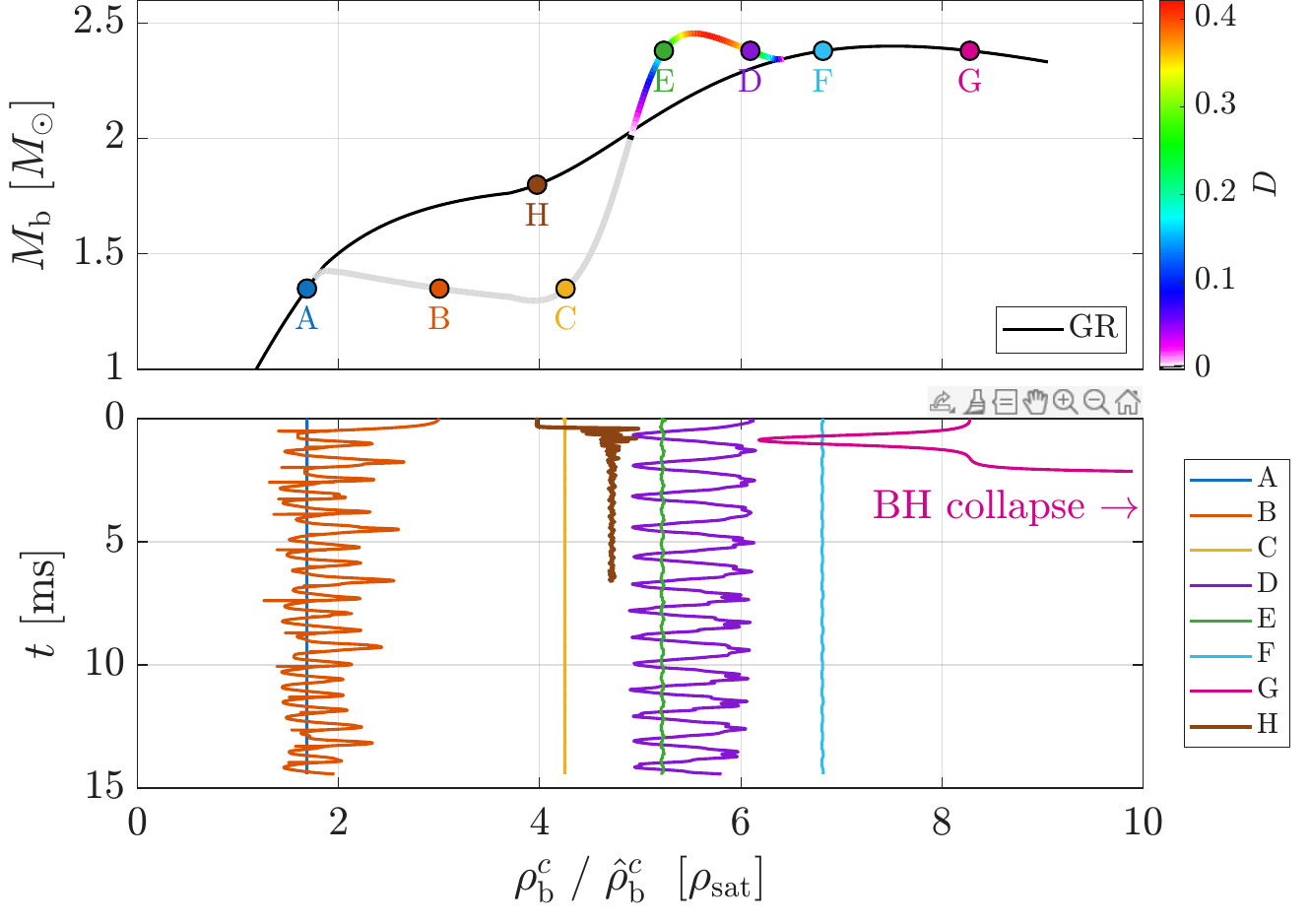}
    \caption{Baryonic mass--number-density relation and dynamical behavior of scalarized neutron-star equilibria in the DEF model for a modified PS EOS. Top: equilibrium sequence in the $M_{\rm b}$--$n_b$ plane. The GR branch is shown in black, while the scalarized branch is color coded by the donutization parameter $D$.
    Points A--C denote representative models with $M_{\rm b}=1.35\,M_\odot$, points D--G denote representative models with $M_{\rm b}=2.38\,M_\odot$, and point H denotes an unstable model with $M_{\rm b}=1.8\,M_\odot$ that undergoes a scalarization.
    Bottom: time evolution of the central baryon density for models A--G, showing oscillatory migration for all unstable configurations except one that collapses to a black hole.
    DEF parameters adopted here are $m_\phi=1.336\times10^{-9}$~eV and $B=2.5\times10^4$.
    }
    \label{fig:PS}
\end{figure}

The hydrodynamic variables are evolved in spherical coordinates with a finite-volume scheme in the reference-metric formalism, following~\cite{Cheong:2020kpv}.
We use these evolutions to test the dynamical stability of scalarized equilibria near the turning points of the baryonic-mass--central-density sequence.
The baryonic mass is the relevant control parameter because it is conserved during the evolution.
In \cref{fig:PS}, we select two triplets near each turning point (one conventional and one reverse) at baryonic masses $M_{\rm b}=1.35\,M_\odot$ and $M_{\rm b}=2.38\,M_\odot$, respectively.
The corresponding evolution of $\hat{\rho}_b^c$ and $\rho_b^c$ (bottom) shows that the intermediate segment between the two stable branches is dynamically unstable, and that perturbed models either migrate toward a stable branch at the same baryonic mass or collapse to a black hole.

\section{S.2.~~~Constraint-enforced Initial Data and Donut Number}
The stationary models used in the main text are constructed in quasi-isotropic coordinates on maximal slices, so the trace of the extrinsic curvature vanishes~\cite{Bonazzola:1993zz,Shibata:2007zzb}.
With axial symmetry, the Hamiltonian constraint for these stationary configurations reduces to~\cite{Shibata:2013pra,Taniguchi:2014fqa,Kuan:2023hrh}
\begin{align}
    0=\mathcal{H}=&R-K_{ij}K^{ij}
    -2\left[8\pi A^4 \rho_{\rm h}+(\pa_k\varphi)( \pa^k\varphi) + 2V\right]
\end{align}
in the Einstein frame, where $\rho_{\rm h}:=e^{\uprho+\upgamma}T^{tt}_{\rm b,J}$ is defined by the Jordan-frame baryonic stress-energy tensor, $R$ is the Ricci scalar of 3-metric, and $K_{ij}K^{ij}$ is  the self-contraction term of extrinsic tensor.
For the initial data used in this work, computed on a domain extending to $10^9$ neutron-star radii, the $L_2$ norm of the Hamiltonian constraint is ${\cal O}(10^{-7})$.
This verifies that the configurations evolved in Sec.~S.1 are constraint-satisfying equilibria, rather than approximate profiles.

We define the donut number $D$ to distinguish ordinary scalarization, where the scalar field fills the core, from genuine donutization, where the scalar support is concentrated in an off-center shell.
The construction uses superlevel sets of the Jordan-frame scalar amplitude and is therefore tied to the scalar geometry rather than to a coordinate location of the peak.
Defining $u(\mathbf{x})= |\tilde{\varphi}(\mathbf{x})| / \max (|\tilde{\varphi}|)$, we consider for each threshold $\eta\in(0,1]$ the connected component of the set $\{\mathbf{x}:u(\mathbf{x})\ge \eta\}$ that contains the global maximum of $u$, where $\mathbf{x}=(r,\,\theta)$.
We denote by $V_{\rm hole}(\eta)$ the volume enclosed by the inner boundary of this component and by $V_{\rm out}(\eta)$ the volume enclosed by its outer boundary.
The corresponding equivalent inner and outer radii are defined by
\begin{align}
    4\pi [R_{\rm in}(\eta)]^3 = \,3V_{\rm hole}(\eta),\,\text{and}\;\;
    4\pi [R_{\rm out}(\eta)]^3 = \,3V_{\rm out}(\eta)\,.
\end{align}
The donut number is designed to measure how persistently the scalar profile forms a shell.
We define it as a geometric, accumulated shell-strength measure,
\begin{align}
    D = \max (|\tilde{\varphi}|) \int_0^1 \, \frac{M_{\rm b}(R_{\rm out})-M_{\rm b}(R_{\rm in})}{R_{\rm out}-R_{\rm in}}\, d\eta\,,
\end{align}
where $R_{\rm out}-R_{\rm in}$ is an effective shell thickness.
At each threshold, the integrand measures the baryonic mass carried by the scalarized shell per unit shell thickness.
The prefactor $\max(|\tilde{\varphi}|)$ prevents a weak off-center fluctuation from being weighted like a strongly scalarized shell.
By construction, $D$ becomes nonzero once the scalar profile departs from triviality, but it remains moderate for centrally filled scalarization because the effective thickness is comparable to the scalarized core size.
When a genuine off-center shell forms, $R_{\rm out}-R_{\rm in}$ can become small while the baryonic mass in the shell remains substantial.
The effect is amplified in the outer core, where the same radial thickness encloses a larger volume through the area factor $4\pi r^2$.
A visualization of how $D$ tracks the formation of the donut-shape is provided in~\cref{fig:map10} of the main text.
The top panel of \cref{fig:PS} also showed the donutization number for the scalarized sequence in the $M_{\rm b}$--$\hat{\rho}_{\rm b}^c$ plane.

\begin{figure}[b]
    \centering
    \includegraphics[width=\columnwidth]{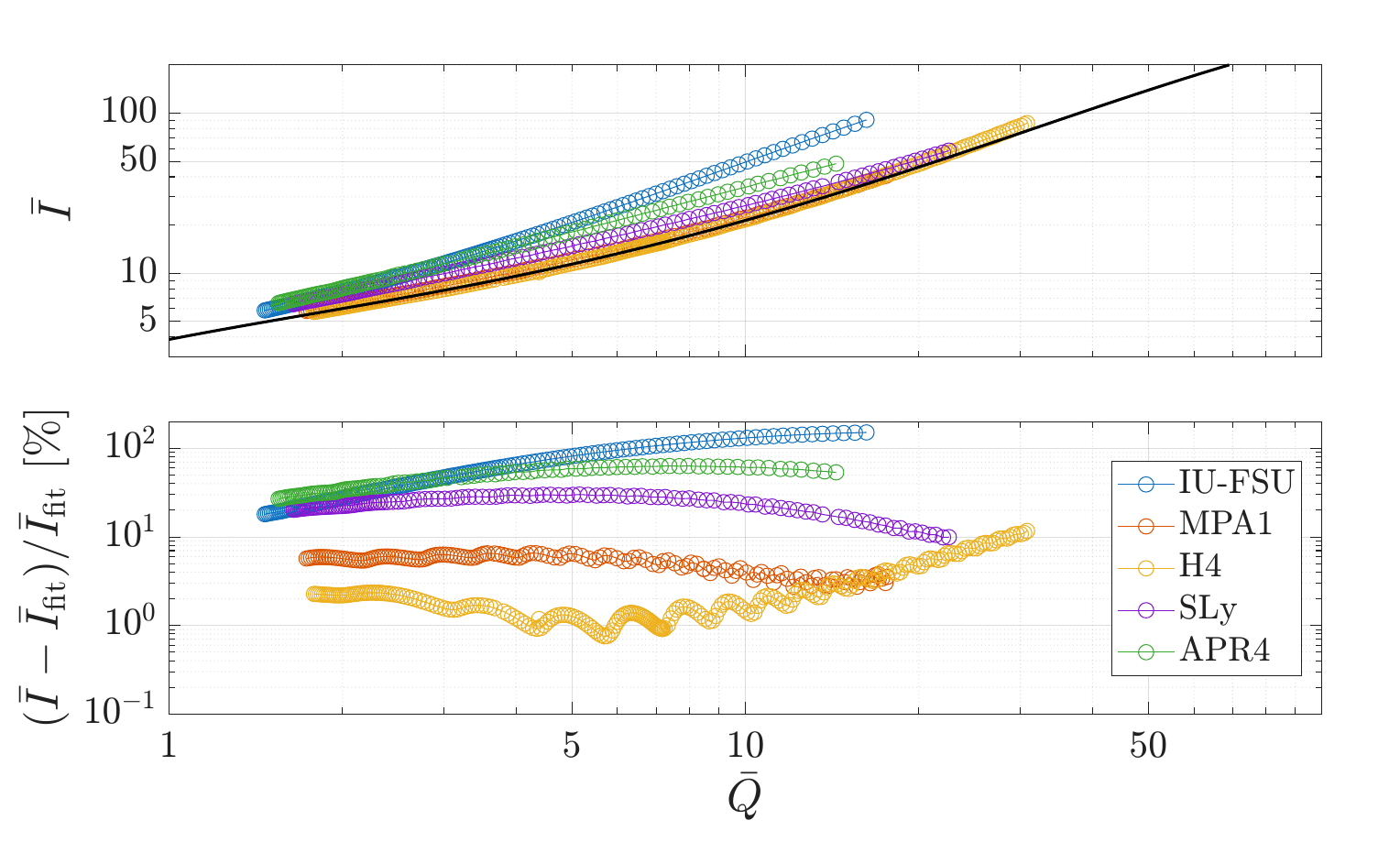}
    \includegraphics[width=\columnwidth]{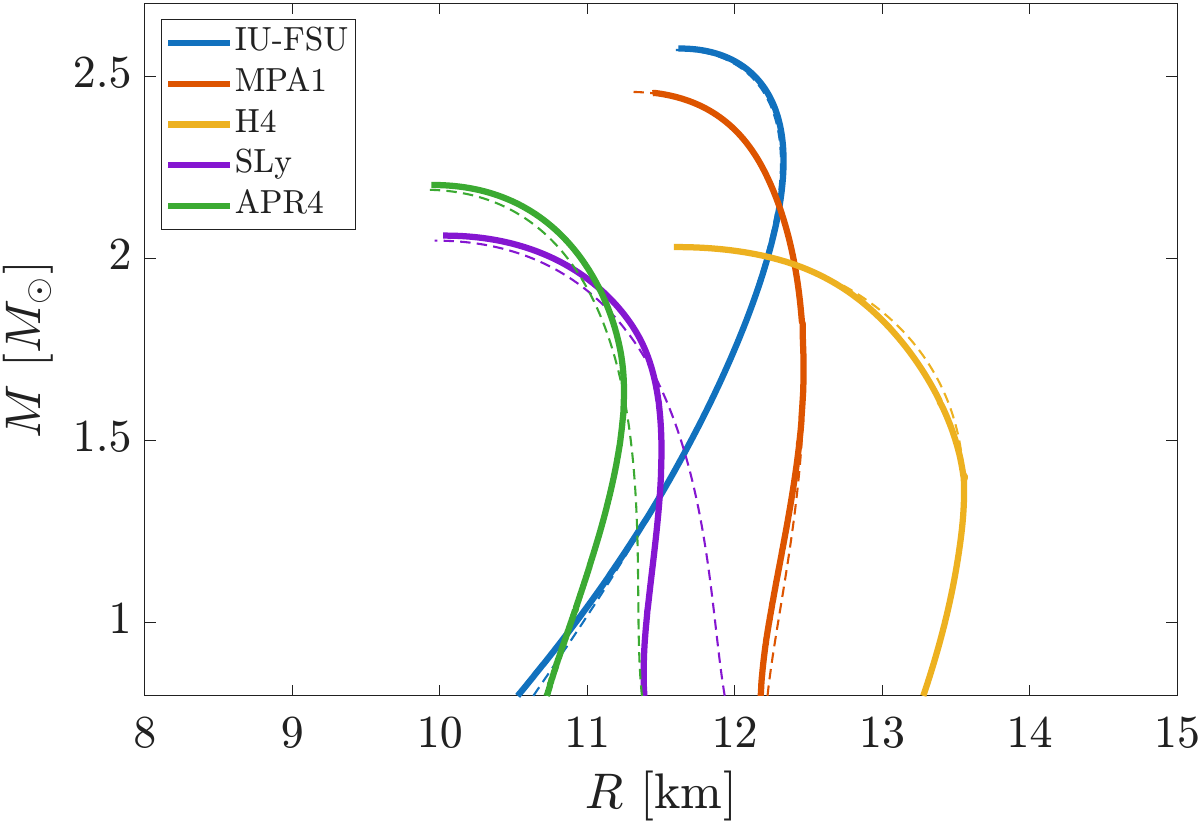}
    \caption{\textbf{Top:} $I$--$Q$ relation for a DEF-type theory (colored) compared with the general-relativistic fitting formula (black curve).
    The lower panel shows the fractional deviation from the fit.
    \textbf{Bottom:} Mass--radius relations corresponding to the model shown above. Equilibria in DEF theory (solid) deviate only slightly from general relativity (dashed) for H4, MPA1, and IU-FSU, but more strongly for APR4 and SLy.
    In both panel, the DEF parameters are $m_\phi=1.336\times10^{-9}$~eV and $B=2.5\times10^4$.}
    \label{fig:IQ}
\end{figure}
\section{S.3.~~~Computation of the moment of inertia}
This section describes the slow-rotation calculation used to obtain the moment of inertia in~\cref{fig:IQ}.
We write the nonrotating background in Schwarzschild, or areal-radius, coordinates as
\begin{align}
    \dd s^2=-e^{\nu} \dd t^2 + e^{\lambda}\dd r_{\rm Sch}^2 + r_{\rm Sch}^2\left(\dd \theta^2 + \sin^2\theta\; \dd \phi^2\right)\,.
\end{align}
The static background equations are
\begin{align}
    \nu'&=\frac{e^{\lambda}-1}{r_{\rm Sch}}
    +r_{\rm Sch} e^{\lambda}  \left(8\pi A^4p - m_\phi^2\varphi^2\right) 
    + r_{\rm Sch}\; \varphi'^2\,,\nn
    \lambda'&=\frac{1-e^{\lambda}}{r_{\rm Sch}}
    +r_{\rm Sch} e^{\lambda}  \left( 8\pi A^4\rho + m_\phi^2\varphi^2 \right)
    + r_{\rm Sch} \; \varphi'^2\,,
\end{align}
where primes denote derivatives with respect to the areal radius $r_{\rm Sch}$.
To first order in the uniform spin frequency $\Omega$, rotation introduces only the azimuthal frame-dragging function $\omega(r)$.
Following the Hartle--Thorne expansion~\cite{Hartle:1967he,Hartle:1968si}, we define $\bar\omega=\Omega-\omega$, the angular velocity of the fluid relative to the local inertial frame, and solve
\begin{align}
    &r_{\rm Sch} \bar\omega'' + 4 \bar\omega'
    =
    \frac{4\pi A^4 r_{\rm Sch}\; \rho h}{1-2m/r_{\rm Sch}}
    \left( r_{\rm Sch} \bar\omega' + 4\bar\omega \right)
    + r_{\rm Sch}^2 \varphi'^2 \bar\omega'
    \nn
    \Rightarrow&
    \frac{1}{r_{\rm Sch}^4} \left(r_{\rm Sch}^4 j \bar\omega' \right)'
    + \frac{4}{r_{\rm Sch}}
    \left( j' + j r_{\rm Sch} \varphi'^2 \right) \bar\omega =0\,,
\end{align}
where $\bar\omega=\Omega-\omega$ is the angular velocity of the fluid element relative to the local inertial frame to linear order of spin rate $\Omega$~\cite{Benhar:2005gi,Yagi:2013awa,Gao:2021uus,Conde-Ocazionez:2025agu}, and $j:= e^{-(\nu+\lambda)/2}=e^{-\nu/2}\sqrt{1-2m/r_{\rm Sch}}$.
The matter term proportional to $A^4\rho h$ is absorbed into $j'$.
Indeed, using the static background equations gives
\begin{align}
    \frac{j'}{j}
    =
    -\frac{4\pi A^4 r_{\rm Sch}\rho h}{1-2m/r_{\rm Sch}}
    -r_{\rm Sch}\varphi'^2\,,
\end{align}
and, thus, the coupling factor $A$ enters implicitly through the background metric combination $j$.

It is useful to rewrite this equation in terms of the quasilocal moment of inertia~\cite{Grigorian:1997Ap},
\begin{align}
    {\cal I}(r_{\rm Sch}) = \frac{\bar\omega' r_{\rm Sch}^4}{2(3\bar\omega+r_{\rm Sch}\bar\omega')}\,,
\end{align}
and so
\begin{align}
    {\cal I}'
    = \frac{3\bar\omega r_{\rm Sch}^3 (4\bar\omega'+r_{\rm Sch}\bar\omega'')}{2(3\bar\omega+r_{\rm Sch}\bar\omega')^2}\,.
\end{align}
Substituting the frame-dragging equation gives an ordinary differential equation for $\mathcal I$~\cite{Hu:2023vsq,Dong:2023vxv} as
\begin{align}
    {\cal I}' &= \frac{8}{3}\pi A^4 r_{\rm Sch}^4 \; \rho h 
    \left( 1 - \frac{5{\cal I}}{2r_{\rm Sch}^3} + \frac{{\cal I}^2}{r_{\rm Sch}^6} \right)
    \left( 1-\frac{2m}{r_{\rm Sch}} \right)^{-1} \nn
    &+ {\cal I} r_{\rm Sch}\left(1-\frac{2{\cal I}}{r_{\rm Sch}^3}\right)
    \varphi'^2
\end{align}
in the Schwarzschild (areal) coordinate and the Einstein frame.
The scalar-gradient term is the direct modification relative to the general-relativistic slow-rotation equation.
To integrate this equation together with the equilibrium models used in the main text, we also evolve the coordinate transformation between areal and isotropic radii,
\begin{align}
    \frac{\dd r_{\rm Sch}}{\dd r}&=\frac{r_{\rm Sch}}{r} \left(1-\frac{2m}{r_{\rm Sch}}\right)^{1/2}\,,\\
    \frac{\dd {\cal I}}{\dd r}&={\cal I}' \frac{\dd r_{\rm Sch}}{\dd r}\,,
\end{align}
where $r$ denotes the isotropic radial coordinate.
The Misner–Sharp mass in the latter coordinates satisfies
\begin{align}
    1-\frac{2m}{r_{\rm Sch}}&:=g^{ab}\left(\pa_ar_{\rm Sch}\right)\left(\pa_br_{\rm Sch}\right)
    =\left( 1+r\frac{\dd \alpha}{\dd r}\right)^2 \nn
    &=\left[ 1+s(1-s)\frac{\dd \alpha}{\dd s} \right]^2
    \,,
\end{align}
with central boundary conditions ${\cal I}(r=0)=0=r_{\rm Sch}(r=0)$.
The asymptotic value of ${\cal I}$ gives the stellar moment of inertia $I$ used in the $I$--$Q$ comparison.

The resulting comparison is shown in~\cref{fig:IQ}.
For the representative massive DEF model used in the Letter, scalar-induced pressure anisotropy produces visible departures from the general-relativistic $I$--$Q$ fitting formula over the astrophysical relevant mass range.
This effect is not simply a restatement of $M$--$R$ changes. 
In fact, we chose specific values of $(B,m_\phi)$ that allow the scalarized version of different baryonic equations of state to nearly reproduce the $M-R$ sequence of general relativity. 
Using various equations of state, such as those with only $npe\mu$ matter (IU-FSU, APR4 \cite{Akmal:1998cf}, SLy \cite{Douchin:2001sv}, and MPA1 \cite{Muther:1987xaa}) and one with hyperonic matter (H4 \cite{Lackey:2005tk}), we find deviations in $I-Q$ scaling up to $100\%$ (although the hyperonic equation of state finds minimal deviations). 
Thus the $I$--$Q$ relation provides an independent diagnostic of the scalarized interior. 

\section{S.4.~~~Donutization dependence on $(B,m_\phi)$}

\begin{figure}[h!]
    \centering
    \includegraphics[width=1\columnwidth]{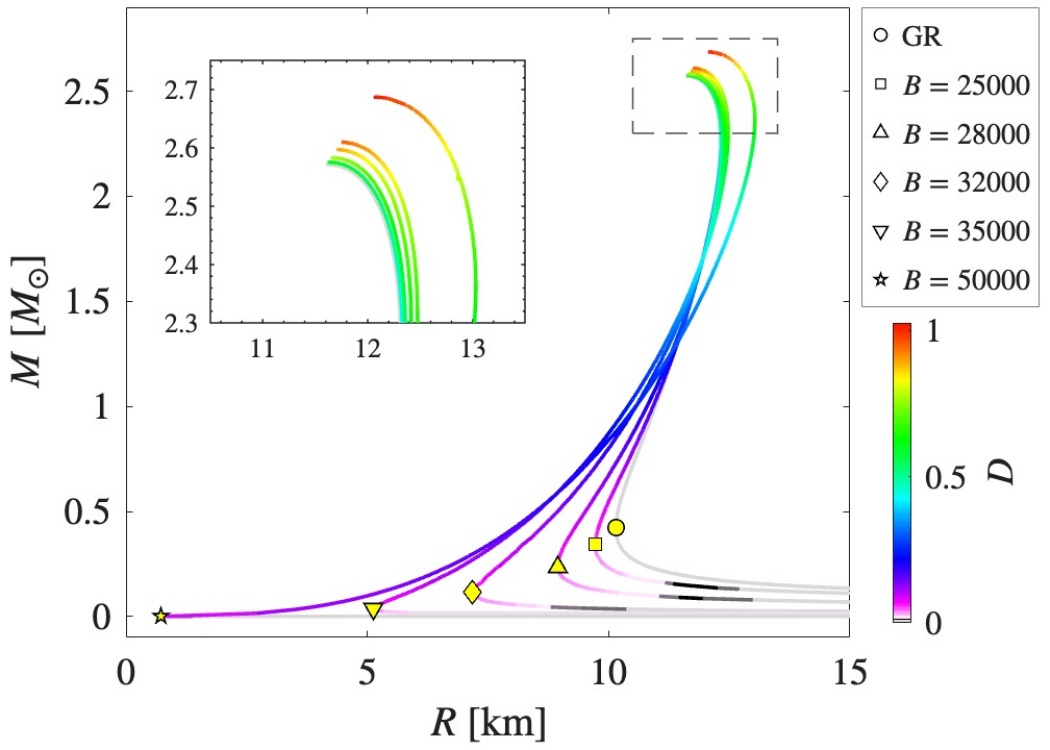}\\
\includegraphics[width=1\columnwidth]{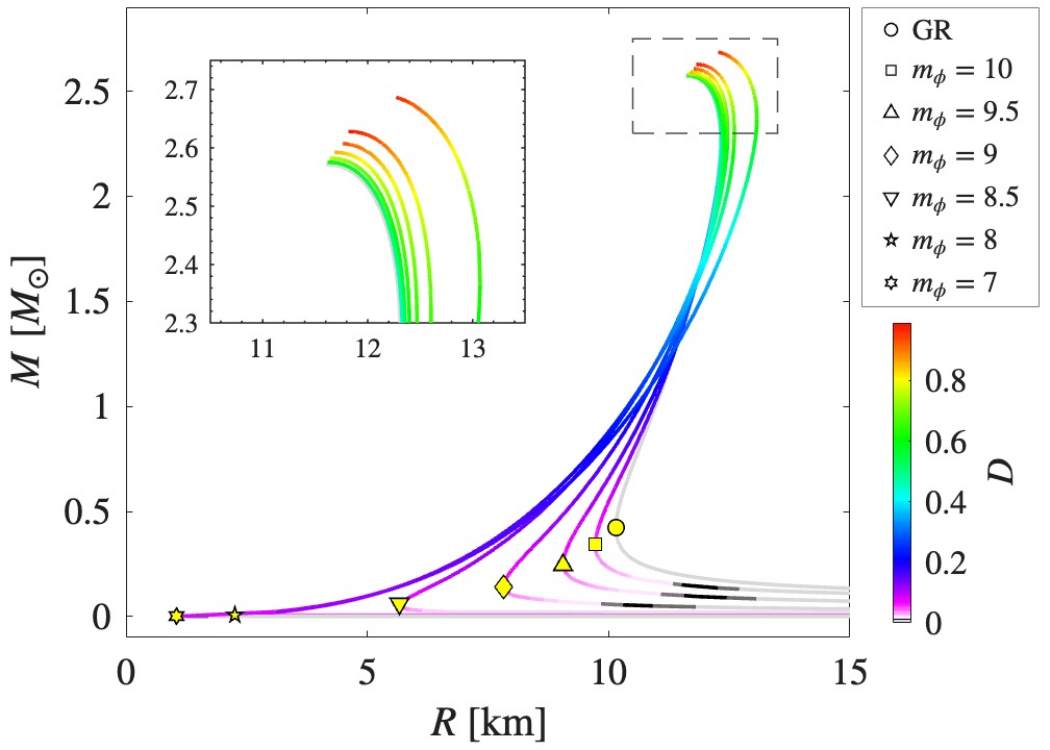}
    \caption{Mass-radius sequences while varying $B$ while holding $m_\phi=1.3\times 10^{-9}$ eV (left) and $m_\phi$ while holding $B=2.4\times10^4$ (right). 
    The lines are color coded by the donutization number, and the different parameters are presented by various markers.
    Legend for $m_\phi$ is given in units of $1.3\times10^{-10}$~eV.
 }
    \label{fig:vary}
\end{figure}

In Fig.\ \ref{fig:vary} we take the IU-FSU baryon equation of state and vary $(B,m_\phi)$ (while holding the other fixed) to understand the interplay between the mimicry of quark star features as well as $D$. 
Across all variations, we find very consistent donutization behavior. Donutization appears close to the kink at low $M$, where the $M$-$R$ curve begins to mimic a quark star.  At these low $M$, $D$ remains small but grows until the maximum mass is reached (the maximum is at $D\sim 0.6-0.9$). The scalarization that most closely mimics the $M$-$R$ sequence of a quark star, leads to the largest $D$. 

In the top of Fig.\ \ref{fig:vary}, we find that, the larger values of $B$ (while holding $m_\phi=1.3\times 10^{-9}$ eV fixed), the stronger the deviations from GR and the more the $M$-$R$ sequence resembles that of a quark star. Large $B$ shifts the maximum mass to higher values and the low $M$ regime has the typical $M\propto R^3$ scaling. At extremely low $M$, we see the $M$-$R$ sequence bends to the right, which is typical due to the presence of a crust. Larger values of $B$ shift the low $M$ bending behavior to lower and lower $M$. 

In the bottom of Fig.\ \ref{fig:vary}, we fix $B=2.4\times 10^4$ and vary $m_\phi$, and find a more subtle effect. There is a critical $m_\phi=7\cdot (1.3\times10^{-10})$~eV value that has a maximal effect in scalarization. Further increases of $m_\phi$ shifts the $M$-$R$ sequence back to the GR curve.  This shift towards GR at heavier $m_\phi$ is also consistent with a smaller donutization number. That being said, even for the largest $m_\phi$ that we plot here, which provides a $M$-$R$ sequence quite close to the GR one, we find significant donutization up to $D\sim 0.6$ at its maximum mass. 

\end{document}